\documentclass{ifacconf}  

\usepackage{amsfonts}
\usepackage{amsmath}
\usepackage{amssymb}
\usepackage{color}
\usepackage[numbers]{natbib}
\usepackage{graphicx}
\usepackage{listings}
\usepackage{xspace}
\usepackage{tikz}
\usepackage{subcaption}
\usetikzlibrary{calc, positioning, fit, arrows, arrows.meta, shapes.geometric, shapes.symbols, shapes.misc, patterns}
\usepackage{algpseudocode}
\usepackage[boxruled,vlined,linesnumbered,onelanguage]{algorithm2e}

\usepackage{xspace}
\usepackage{tikz}
\usepackage{subcaption}
\usetikzlibrary{calc, positioning, fit, arrows, arrows.meta, shapes.geometric, shapes.symbols, shapes.misc, patterns}

\newcommand{\R}{{\rm I\!R}}

\def\mathscr#1{{\cal #1}}
\newtheorem{theorem}{Theorem}

\newtheorem{example}{Example}

\newcommand{\0}{\mathbf{0}}

\newtheorem{definition}{Definition}

\def\send#1#2{\stackrel{#1}{\hbox to #2{\rightarrowfill}}}
\def\-{\!\!\!\!\!-}

\def\R{{\rm I\!R}} 

\newcounter{seqn}[equation]
\def\theseqn{\arabic{equation}\alph{seqn}}

\def\endseqn{\eqno \@seqnnum
$$\ignorespaces}
\def\@seqnnum{(\theseqn)}

\newskip\mcentering \mcentering=0pt plus 1000pt minus 1000pt

\def\meqalignno#1{
\halign to\displaywidth{
    \hbox to 0pt{\kern\displaywidth\llap{$##$}\hss}\tabskip=\mcentering
    &\hfil$\displaystyle{##}$\tabskip=\mcentering
   &&$\displaystyle{{}##}$\hfil\tabskip=\mcentering
    \crcr
    #1\crcr}}

\def\dspace{\multiply\normalbaselineskip 150
		  \divide\normalbaselineskip 100 \normalbaselines
		  \csname @@normalbaselineskip\endcsname\normalbaselineskip}
\def\sspace{\multiply\normalbaselineskip 200
		 \divide\normalbaselineskip 300 \normalbaselines
		 \csname @@normalbaselineskip\endcsname\normalbaselineskip}
\def\sdspace{\multiply\normalbaselineskip 160
		 \divide\normalbaselineskip 150 \normalbaselines
		 \csname @@normalbaselineskip\endcsname\normalbaselineskip}

\def\@{\tilde}

\def\3dot#1{\buildrel\textstyle...\over#1}

\newcommand{\drawnodex}[6]{
	\path (#1) node[draw,thick,circle,minimum width=1em,minimum height=1em,inner sep=0.2ex] (#2) {} node[inner sep=0.2ex,fill=white,#3=0 of #2] {#4} node[inner sep=0.2ex,fill=white,#5=0 of #2] {#6};
}

\newcommand{\drawfoex}[5][0.5]{
	\path (#2) -- (#3) node(_1)[allow upside down,sloped,above,minimum height=1.2em,anchor=center,pos=0.3] {};
	\path (#2) -- (#3) node(_2)[allow upside down,sloped,above,minimum height=1.2em,anchor=center,pos=0.7] {};
	\draw[-stealth,ultra thick,red] (#2) .. controls (_1.north) and (_2.north) .. (#3) node[inner sep=0.2ex,fill=white,sloped,pos=#1] {#4};
	\draw[-stealth,ultra thick,red] (#3) .. controls (_2.south) and (_1.south) .. (#2) node[inner sep=0.2ex,fill=white,sloped,pos=#1] {#5};
}

\begin{document}

\begin{frontmatter}

\title{The Power Allocation Game on A Network: Computation Issue\thanksref{footnoteinfo}}


\thanks[footnoteinfo]{This work was supported by National Science Foundation grant n.1607101.00 and US Air Force grant n. FA9550-16-1-0290.}

\thanks[footnoteinfo]{The authors thank Thomas Pogge for valuable suggestions on the case study.}



\author[First]{Yuke Li} 
\author[Second]{Jiahua Yue} 
\author[Third]{Fengjiao Liu}
\author[Fourth]{A. Stephen Morse}

\address[First]{Department of Political Science, Yale University, 06511, USA (e-mail: yuke.li@yale.edu).}
\address[Second]{Department of Political Science, Yale University, 06511, USA (e-mail: jiahua.yue@yale.edu)}
\address[Third]{Department of Electrical Engineering, Yale University, 06511, USA (e-mail: fengjiao.liu@yale.edu)}
\address[Fourth]{Department of Electrical Engineering, Yale University, 06511, USA (e-mail: as.morse@yale.edu)}


\begin{abstract}

In this paper two algorithms with the goal of generating the equilibrium set of the power allocation game first developed in \cite{allocation} are proposed. Based on the first algorithm, the geometric property of the pure strategy Nash equilibrium set will be proven to be a collection of convex polytopes. The second, simulation-based, algorithm is developed to overcome the shortcoming of the first algorithm in terms of generating the equilibrium set efficiently and then making policy-relevant predictions based on the set. The second algorithm will be usefully applied to a real-world case study, which draws on the current crisis between North Korea and certain key players including the US and China.  

\end{abstract}
\begin{keyword} 
Nash equilibrium computation, total order, best response, constraint satisfaction, convex polytope
\end{keyword}

\end{frontmatter}

\section{Introduction}

A framework of the static power allocation game (abbreviated as PAG) was first developed in \citep{allocation}, where agents simultaneously allocate the total power to their relations subject to both the total power constraints and their preferences for the power allocation outcomes 
and reach a Nash equilibrium if and only if no one will strictly benefit in terms of unilateral deviation.  This game was developed to specifically suit the context of strategic interactions in a networked, adversarial environment. \citep{allocation} presents the proof of the existence of pure strategy Nash equilibrium for any parametric variation of the power allocation game. 

Obviously, it is the states of the countries predicted by the equilibria rather than the equilibria themselves that should be of interest. This is because an infinite number of equilibria can predict the same combination of states of all countries; besides, the states provide valuable information as to whether a country can survive in equilibrium. Therefore, in \citep{allocation}, the idea of {\em  equivalence relations} on the set of equilibria $U^*$ for the game was introduced. Two equilibria are {\em equivalent} if and only if they yield the same predictions for the entities' states. In general, multiple equilibrium equivalence classes exist in many parametric variations of the game. The existence of multiple equilibrium equivalence classes makes a probabilistic prediction of the power allocation outcomes in different networked international environments necessary. 

However, no discussion on finding \emph{all} possible pure strategy Nash equilibria or the equilibrium equivalence classes was given yet. The problem considered in this paper is to execute a search of all possible pure strategy Nash equilibria and then derive all the equilibrium equivalence classes by partitioning the set based on the equivalence relations. Such a predictive study of the power allocation outcomes in different networked, adversarial environments is quite necessary. By making predictions for power allocation outcomes for each possible networked environment, it will help a decision maker to rationally decide the kind of changes that ought to be made to the environment in its favor. 


The intellectual background for this research problem is the literature on Nash equilibrium computations. Ever since Nash's famous paper in 1950, the issue of Nash equilibrium computations has traditionally been focused on the computation of mixed strategy Nash equilibrium in two-player or n-player finite games \citep{lemke1964equilibrium}. The main idea is to study the equilibrium computation problem first as a fix point problem that
is easy to solve, and then continuously transforms it into the actual problem of interest by ``pivoting'' to new fix points \citep{nisan2007algorithmic, mckelvey1996computation}. Equilibrium computations of a similar flavor include the Harsanyi-Selten adjustment process \citep{goldberg2013complexity}, though the adjustment process in \citep{harsanyi1988general} was originally developed to suit the problem of equilibrium selection rather than the problem of equilibrium computations. Closely related papers on fixed point computations such as the Scarf's simplicial subdivision algorithm  \citep{scarf1973computation} include \citep{herings2010homotopy, eaves1972homotopies, scarf1973computation}, and a collection of relevant work is discussed in \citep{karamardian2014fixed}. 

Existing methods may be not suitable for addressing this task for two reasons. First, the power allocation game is an $n$-player game with an infinite strategy space; meanwhile, existing methods mostly apply to games with a finite strategy space. Though a finite-approximation of the strategy space is possible, it will not be the focus of this project. Second, by the nature of the power allocation game itself, a pure strategy Nash equilibrium is easier to be described and more realistic than a mixed strategy Nash equilibrium. 

Following a brief review of the power allocation game, a combinatorial algorithm will be proposed. The input of the algorithm will be a collection of the parameters which defines a power allocation game, and the output of the algorithm will be the set of all possible pure strategy Nash equilibria. The algorithm proceeds by checking each possible candidate equilibrium equivalence class and determining its validity. Based on the algorithm, the geometric property of the equilibria set will be established to be a collection of convex polytopes. A simulation-based algorithm for efficiently generating the equilibrium set of the power allocation game will then be proposed, whose predictive capability for events will be illustrated with a case study drawing on one current real world event.




\section{The PAG}

\subsection{Basic Idea} By the  {\em power allocation game} or PAG is meant a distributed resource allocation game between $n$ countries with labels in $\mathbf{n}= \{1,2,\ldots,n\}$\citep{allocation}. The game is formulated on a simple, undirected,  signed graph $\mathbb{G}$ called ``an environment graph'' \citep{survival} whose $n$ vertices correspond to the countries and whose $m$ edges represent relationships between countries. An edge between distinct vertices $i$ and $j$, denoted by $(i,j)$, is labeled with a plus sign if countries $i$ an $j$ are friends and with a minus sign if countries $i$ and $j$ are adversaries.
For each $i\in\mathbf{n}$, $\mathscr{F}_i$ and $\mathscr{A}_i$ denote the sets of labels of country $i$'s friends and adversaries respectively; it is assumed that $i\in\mathscr{F}_i$ and that $\mathscr{F}_i$ and $\mathscr{A}_i$ are disjoint sets.
Each country $i$ possesses a nonnegative quantity $p_i$ called the {\em total power} of country $i$. An allocation of this power or {\em strategy} is a nonnegative $n\times 1$ row vector $u_i$ whose $j$ component $u_{ij}$ is that part of $p_i$ which country $i$ allocates under the strategy to either support country $j$ if $j\in\mathscr{F}_i$ or to demise country $j$ if $j\in\mathscr{A}_i$; accordingly $u_{ij}= 0$ if $j\not\in\mathscr{F}_i\cup\mathscr{A}_i$ and $u_{i1}+u_{i2}+\cdots +u_{in} = p_i$. The goal of the game is for each country to choose a strategy which contributes to the demise of all of its adversaries and to the support of all of its friends.

Each set of country strategies $\{u_i,\;i\in\mathbf{n}\}$ determines an
 $n\times n$ matrix  $U$ whose $i$th row is $u_i$. Thus $U = [u_{ij}]_{n\times n}$ is a nonnegative matrix such that, for each $i\in\mathbf{n}$, $u_{i1}+u_{i2}+\cdots +u_{in} = p_i$. Any such matrix is called a {\em strategy matrix}  and $\mathscr{U}$ is the set of all $n\times n$ strategy matrices.

\subsection{Multi-front Pursuit of Survival}

How countries allocate the power in the support of \emph{the survival} of its friends and the demise of that of its adversaries is studied in \citep{survival} and \citep{balancing} in line with the fundamental assumptions about countries' behavior in classical international relations theory.\citep{waltz1979theory} The following additional formulations are offered: 

Each strategy matrix $U$ determines for each $i\in\mathbf{n}$, the {\em total support} $\sigma_i(U)$ of
 country $i$ and the {\em total threat}  $\tau_i(U)$ against country $i$.  Here
 $\sigma_i:\mathscr{U}\rightarrow\R$ and $\tau_i:\mathscr{U}\rightarrow \R$ are non-negative
 valued maps defined by
$U\longmapsto \sum_{j\in\mathscr{F}_i}u_{ji} +\sum_{j\in\mathscr{A}_i}u_{ij}$
 and $U\longmapsto \sum_{j\in\mathscr{A}_i}u_{ji}$
  respectively. Thus country $i$'s total support is the sum of the amounts of power
   each of country $i$'s friends
  allocate to its support plus the sum of the amounts of power country $i$ allocates  to the
  destruction of  all of its adversaries. Country $i$'s total threat, on the other hand, is the sum of the amounts of power
  country $i$'s adversaries allocate to its destruction. A \emph{state function} $x: \mathcal{U} \rightarrow \mathcal{X}$ maps a power allocation matrix to a $n$ vector $x(U)$ (defined as \emph{the state vector}), where the $i$-th element $x_i(U)$ is country $i$'s \emph{state}, 
     $x_i(U) = $ safe
    if $\sigma_i(U)>\tau_i(U)$, $x_i(U)=$ precarious if $\sigma_i(U)=\tau_i(U)$, or
    $x_i(U) = $ unsafe if $\sigma_i(U)<\tau_i(U)$. $\mathcal{X} = \{\text{safe}, \text{precarious}, \text{unsafe}\}^{n}$ is the \emph{state space}.

In playing the PAG, countries select individual strategies in accordance with certain weak and/or strong preferences.
A sufficient set of conditions for  country $i$ to {\em weakly prefer} strategy
matrix $V\in\mathscr{U}$ over strategy matrix $U\in\mathscr{U}$ are as follows
\begin{enumerate}
\item  For all $j\in\mathscr{F}_i$
 either $x_j(V)\in$ \{safe, precarious\}, or $x_j(U)\in$ \{unsafe\}, or both.
 \item For all $j\in\mathscr{A}_i$ either $x_j(V)\in $ \{unsafe, precarious\}, or $x_j(U)\in$ \{safe\}, or both.
\end{enumerate}
Weak preference  by country $i$ of  $V$ over $U$ is denoted by $U \preceq_{i} V$.

Meanwhile, a sufficient condition for country
 $i$ to be {\em indifferent} to the choice between $V$ and $U$ is  that $x_i(U)=x_j(V)$ for all $j\in\mathscr{F}_i\cup\mathscr{A}_i$.  This is denoted by $V\sim_{i} U$.

Finally, a sufficient condition for country $i$  to {\em strongly prefer} $V$ over $U$ is that $x_i(V)$ be a  safe or precarious  state and $x_i(U)$ be an unsafe state. Strong preference by country $i$ of $V$ over $U$ is denoted by $U \prec_{i} V$.

The two sufficient conditions were stated as two axioms in \citep{allocation} and \citep{survival}) and not definitions per se of the weak preference or strict preference. They determine \emph{a partial order} of the power allocation matrices in $\mathscr{U}$, and the two axioms are consistent.  A valid total order that satisfy the two axioms exists. 

The Nash equilibrium concept is naturally employed to make predictions for the PAG. Let country $i$'s deviation from the power allocation matrix $U$ be a nonnegative-valued $1\times n$ row vector $d_{i} \in \mathbb{R}^{1 \times n}$ such that $u_{i} + d_{i}$ is a valid strategy that satisfies the total power constraint for country $i$. The deviation set $\mathscr{\delta}_{i}$ is the set of all possible deviations of country $i$ from the power allocation matrix $U$. In the context of a PAG, a power allocation matrix $U$ is a pure strategy Nash Equilibrium if no unilateral deviation in strategy by any single country $i$ is profitable for $i$, that is, 
\begin{equation*}
U + e_{i} d_{i} \preceq U, \;\;\;\;\; {\rm for\; all} \;\;\;\;\; d_{i} \in \mathscr{\delta}_{i},
\end{equation*} 
where $e_{i}$ is an $n \times 1$ unit vector whose elements are $0$ but the $i$-th coordinate which is $1$.

Denote by $\mathcal{U}^{*}$ the set of pure strategy Nash equilibria. Call 
 $U\in\mathcal{U}^*$  {\em equilibrium equivalent} to $V\in\mathcal{U}^*$ if and only if $x(U) = x(V)$. The relation ``equilibrium equivalence'' is an equivalence relation on $\mathcal{U}^{*}$.
Let $[U]_*$ be the \emph{equilibrium equivalence class} of $U \in \mathcal{U}^{*}$. Obviously, the total number of equilibrium equivalence classes is at most $3^{n}$, which in turn is the cardinality of the co-domain of $x$, $\mathcal{X}$.


\section{The Algorithm}

A combinatorial algorithm is developed in this section to solve the problem of generating the equilibrium set $\mathscr{U}^*$ and then the equilibrium classes of the PAG. The general goal of the algorithm is to derive the algebraic representations of the equilibrium set by checking whether the intersections of countries' best responses under each possible equilibrium equivalence class defined by the relevant state vector is nonempty within the feasible regions of the strategy space. 

\subsubsection*{Input of the Algorithm.} The input of the algorithm consists of the countries set, $\mathbf{n}$, the total power vector $p$, the friend set, the adversary set and the preference order of each country $i$, $\mathcal{A}_{i}, \mathcal{F}_{i}, \preccurlyeq_{i}; i \in \mathbf{n}$ which are valid total orders satisfying the two axioms (see the discussion in \cite{paradox} and \cite{survival}). 

Next the algorithm derives three sets of constraints an equilibrium class must satisfy: each country's \emph{total power constraints}, \emph{state constraint}, and \emph{best response constraint}. 

\subsubsection*{Total Power Constraints}

For each country $i$, let its total order of the $3^{n}$ elements in $\mathcal{X}$ of $i$ be $$X_1\succeq_{i} X_2 \cdots \succeq_{i} X_{k} \succcurlyeq_{i} \cdots \succeq_{i}  X_{3^{n}}.$$ 

Given the $k$-th state vector, $X_{k}$, which defines the $k$-th \emph{candidate equilibrium equivalence class}, obtain $i$'s \emph{total power constraint}, $\mathcal{P}_{i}(k)$ defined as the intersection of constraints (i.e., linear inequalities) for each dimension of allocation by $i$ that have to hold in equilibrium. These linear inequalities give the upper and lower bounds for each dimension of allocation by $i$ for any power allocation matrix $U$.

$$\mathcal{P}_{i}(k) := \bigcap_{j \in \mathscr{F}_{i}\cup \mathscr{A}_{i}} (u_{ij} \geq 0) \bigcap (\sum_{j \in \mathcal{F}_{i} \cup \mathcal{A}_{i}}u_{ij} = p_{i})$$


\subsubsection*{State Constraints}

For the $k$-th \emph{candidate equilibrium equivalence class}, obtain each $i$'s \emph{state constraint}, $$\mathcal{S}_{i}(k) := \begin{cases}
(\sigma_{i}(U) > \tau_{i}(U)) & x_{i}(U) = \text{safe} \\
(\sigma_{i}(U) = \tau_{i}(U)) & x_{i}(U) = \text{precarious} \\
(\sigma_{i}(U) < \tau_{i}(U)) & x_{i}(U) = \text{unsafe}  
\end{cases}$$ defined as the intersections of the constraints for the allocations in $i$'s total support $\sigma_{i}$ and total threat $\tau_{i}$ for the relevant equilibrium equivalence class to hold.





\subsubsection*{Best Response Constraints} 

Based on country $i$'s total power constraints and the state constraints, now obtain each country $i$'s \emph{best response constraint}. For the $k$-th best response constraint $\text{Br}_{i}(k)$, it is obtained by intersecting countries' relevant state and total power constraints under the $k$-th candidate equilibrium equivalence class, $$\text{Br}_{i}(k) = \bigcap_{j\in \mathscr{A}_{i} \cup \mathscr{F}_{i}}(\mathcal{S}_{j}(k) \cap \mathcal{P}_{j}(k)).$$ 

The the right-hand side of the above equation can be expanded into a Conjunctive Normal Form (CNF) of $T_i \geq 0$ disjunctive clauses. 

$$\text{CNF}(\bigcap_{j\in \mathscr{A}_{i} \cup \mathscr{F}_{i}}\mathcal{S}_{j}(k) \cap \mathcal{P}_{j}(k)) = \bigcap_{1 \leq t_{i} \leq T_{i}} \text{Clause}_{t_i}.$$  

The $t_{i}$-th clause ($0 \leq t_{i} \leq T_{i}$) is said to be \emph{a clause for $i$'s own allocations under the $k$-th candidate equilibrium equivalence class}, denoted as $\text{Strat}_{t_{i}}(k)$, if and only if it contains only the constraints for allocations by $i$. 

Otherwise, it is said to be \emph{a clause for others' allocations in the $k$-th candidate equilibrium equivalence class}, denoted as $\text{Cond}_{t_{i}}(k)$. 

It is possible to further expand $\text{Strat}_{t_{i}}(k)$ or $\text{Cond}_{t_{i}}(k)$, for instance, expanding each subclause of them into a (smaller) conjunctive normal form.

A \emph{non-deviation condition} for $\bigcap_{i \in \mathbf{n}}\text{Br}_{i}(k)$ has to be satisfied such that for any $i \in \mathbf{n}$, $i$ cannot deviate to a preferred equilibrium equivalence class to the $k$-th one by unilaterally changing its strategy. Let $\bigcap_{i \in \mathbf{n}}\text{Br}_{i}(k)$ that satisfies the non-deviation condition as $\bigcap_{i \in \mathbf{n}}\text{Br}_{i}(k)^*$.

The pure strategy Nash equilibrium set $\mathscr{U}^*$ is the union of the intersections of all countries' best response constraints for every candidate equilibrium equivalence class, $$\mathscr{U}^*:= \bigcup_{1 \leq k \leq 3^n} \Big(\bigcap_{i \in \mathbf{n}}\text{Br}_{i}(k)^*\Big).$$

The output of the algorithm is each equilibrium equivalence class in the algebraic representation. The algebraic representation is generated with the aim of computing the volume of the class, which can then used for calculating, for instance, the likelihood of survival for a given country in the given power allocation game. (where \emph{survival} is defined by being in the state of safe or precarious in \cite{survival}.) However, obtaining the algebraic representation of the classes using the algorithm can be difficult, even for the simple example in Section 4. This is not surprising because the problem of generating the equilibrium equivalence classes of the power allocation game is a \emph{constraint satisfaction problem} (CSP), which can be be of high complexity.

\section{Main results}\label{pr} 

This section contains a proof of the main result of this paper, which is that 
 the equilibrium equivalence classes are geometrically \emph{convex polytopes}.

\begin{definition}[Half-space representation of Convex Polytopes]A half-space representation of a polytope is $P = \{x \in \mathbb{R}^{n}:  Ax \leq b\}$, where $A \in \mathbb{R}^{m \times n}$, $b \in \mathbb{R}^{m \times 1}$, with $m, n \in \mathbb{Z}$ ($m > n \geq 1$).
\end{definition}

\begin{theorem}[Strategy Space as Polytope]
\label{prop/polytope}
Given the power allocation game, the strategy space is a $2m$-dimensional convex polytope, where $m$ is the number of pairs of distinct countries who are friends or adversaries.   
\end{theorem}

\begin{pf} Let the set of allocations between any pair of distinct countries who are friends or adversaries be $$\{u_{ij}: j \in \mathcal{F}_{i} \cup \mathcal{A}_{i} - \{i\}, i \in \mathbf{n}\} = \bigcup_{i \in \mathbf{n}}\{u_{ij}: j \in \mathcal{F}_{i} \cup \mathcal{A}_{i} - \{i\}\}$$ whose cardinality is $2m$. 

Label all allocations in $\{u_{ij}: j \in \mathcal{F}_{i} \cup \mathcal{A}_{i} - \{i\}, i \in \mathbf{n}\}$; the labeling set is $\mathbf{z} =\{1,2,\ldots,2m\}$.  $\mathbf{z} = \bigcup_{i \in \mathbf{n}}\mathbf{z}_{i}$, where $\mathbf{z}_{i}$ is the labeling set for the set of $i$'s allocations in $\{u_{ij}: j \in \mathcal{F}_{i} \cup \mathcal{A}_{i} - \{i\}, i \in \mathbf{n}\}$.



For each $n \times n$ allocation matrix $U$, a $2m$-dimensional vector $\hat{u}$ can be constructed such that $\forall i, j \in \mathbf{n}$ such that $i \neq j$, the $k$-entry, $\hat{u}_{k}$ is equal to the $k$-allocation in $\{u_{ij}: j \in \mathcal{F}_{i} \cup \mathcal{A}_{i} - \{i\}, i \in \mathbf{n}\}$ ($1 \leq k \leq 2m$), with the idea being that the total power constraint, the number of independent nonzero entries in $U$, i.e., the number of independent power allocations is $2m$. The projection $\pi: U \mapsto \hat{u}$ is bijective. As $\pi$ is a bijection, $\hat{u}$ and $U$ can be used interchangeably.

Let $A = [a_{ij}]_{n \times 2m}$ be a real matrix, where the $i$-th row vector, $a_{i}$, is defined such that $a_{iq} = 1$ if and only if $q \in \mathbf{z}_{i}$, and $0$ otherwise. $I$ is an identity matrix, $p = [p_{i}]_{n \times 1}$, and $0= [0]_{n \times 1}$.

The following expression yields both the total power constraints and the nonnegative requirements for all the allocations. $$P=\left\{\hat{u} \in \mathbb{R}^{2m \times 1}: \left[\begin{smallmatrix}A\\- I\end{smallmatrix}\right] \hat{u} \leq\left[\begin{smallmatrix}p\\0\end{smallmatrix}\right]\right\}.$$ 
which is in the half-space representation of the strategy space.

Therefore, the strategy space of the power allocation game is a $2m$-dimensional convex polytope.  $\square$

 \end{pf}
 


\begin{theorem} [Equilibrium Equivalence Class as Polytope]
\label{prop/polytope1}
Each equilibrium equivalence class of the power allocation game is a convex polytope with at most $2m$ dimensions. The equilibrium set $\mathcal{U}^*$ is a collection of convex polytopes. 
\end{theorem}

\begin{pf} In the algorithm in Section 3, if the non-deviation condition is not satisfied for the $k$-th candidate equilibrium equivalence class, the class will have been ruled out in the equilibrium set. If it does for the $k$-th equilibrium equivalence class, $\bigcap_{i \in \mathbf{n}}\text{Br}_{i}(k)^*$ is a finite intersection of simple linear inequalities of countries' allocations for this class. 

Since each linear inequality represents a half-space, each candidate equilibrium equivalence class is either empty or a nonempty convex polytope. 

Therefore, the equilibrium set $\mathcal{U}^*$ is a union of convex polytopes. $\square$

\end{pf}


\begin{example}
Consider a simple three-player conflict scenario, where the main parameters of the power allocation game are the following:

\begin{enumerate}
\item The countries set: $\mathbf{n} = \{1, 2, 3\}$.
\item Countries' power: $p = [5~5~9]$.
\item The relations: $\mathcal{A}_1 = \{2\}$, $\mathcal{A}_2 = \{1,3\}$, and $\mathcal{A}_3 = \{2\}$. $\mathcal{F}_1 = \mathcal{F}_1 = \mathcal{F}_1 = \emptyset$. 
\item Each country has a total order of all possible state vectors as the following: 

\end{enumerate}

\begin{figure}[htbp]
\centering
				
\begin{subfigure}[t]{0.45\linewidth}
	\centering
	\begin{tikzpicture}[x=4em,y=-4em]
		\drawnodex{0,0}{v1}{below}{$v_{1}$}{above}{$5-a$}
		\drawnodex{1.5,0}{v2}{below}{$v_{2}$}{above}{$9-b-c$}
		\drawnodex{3,0}{v3}{below}{$v_{3}$}{above}{$5-d$}
				
		\drawfoex{v1}{v2}{$a$}{$b$}
		\drawfoex{v2}{v3}{$c$}{$d$}
	\end{tikzpicture}
	\caption{Allocations}
	\label{ile2}
\end{subfigure}
\label{ile}

\end{figure}

With the role of the ``precarious'' state in countries' preference orders (as stated in the axioms in \cite{allocation}), only $2^n$ instead of $3^n$ state vectors need to be strictly ordered. Accordingly, the following simplications will be used. For country $i$, denote country $j$'s state to be 1 if $j$ is $i$'s friend and is safe/precarious or $j$ is $i$'s adversary and is unsafe/precarious, and to be 0 if and only if $j$ is $i$'s friend and is unsafe or $j$ is $i$'s adversary and is safe. If $i$ and $j$ have no relations, $j$'s state is $1$ if it is safe/precarious, and $0$ if it is unsafe. 

Assume for country 1, its total order of the state vectors is: 
\begin{align*}[1, 1, 1] &\sim_{1} [1, 1, 0] \succeq_{1} \\ [1, 0, 1]  &\sim_{1} [1, 0, 0] \succeq_{1} \\ [0, 1, 0]  &\sim_{1} [0, 1, 1] \sim_{1} [0, 0, 1] \sim_{1} [0, 0, 0] \end{align*}

Assume for country 2, its total order of the state vectors is: 
\begin{align*} [1, 1, 1] &\succ_{2} [0, 1, 1] \succ_{2} [1, 1, 0] \succ_{2} [0, 1, 0]  \succ_{2}  \\ [1, 0, 1] &\succ_{2} [0, 0, 1] 
	\sim_{2} [1, 0, 0] \succ_{2} [0, 0, 0]. \end{align*}
	
Assume for country 3,  its total order of the state vectors is: 
\begin{align*}[1, 1, 1] &\sim_{3} [0, 1, 1]   \succeq_{3} \\ [1, 0, 1] &\sim_{3} [0, 0, 1] \succeq_{3} \\ [1, 1, 0] &\sim_{3} [0, 1, 0]
	\sim_{3}  [1, 0, 0] \sim_{3} [0, 0, 0].\end{align*}

The nonempty candidate equilibrium equivalence classes are respectively $[0, 0, 1]$ and $[1, 0, 0]$, represented respectively in conjunctive normal form as \begin{align*}(4 \leq a)&～\bigcap～ (9 - a \leq d) \bigcap (a \leq 5) \bigcap (b \leq 4)  \bigcap \\(d \leq 5)& \bigcap (0 \leq b) \bigcap (5 \leq c) \bigcap( c \leq 9-b)\end{align*} and 

\begin{align*}&(4 \leq a) \bigcap (9-a \leq d) \bigcap (a \leq 5) \bigcap (b \leq 9) \bigcap\\&(c \leq 9-b)\bigcap (0 \leq c) \bigcap (5 \leq b) \bigcap (d \leq 5).\end{align*}

The reason for the existence of the two equilibrium equivalence classes is largely because of the indifference of country $2$ between the state vectors $[0, 0, 1]$ and $[1, 0, 0]$. Had country $2$ preferred a vector over the other, there will only exist a single equilibrium equivalence class predicted by the preferred vector. 
    
    \end{example}

\section{Making Predictions Using Simulations}
    
The concluding remarks in Section 3 and the example in Section 4 prompt a study of how to generate the equilibrium equivalence classes of the power allocation game efficiently and then use them for making predictions. This study can be undertaken using simulations, and the following algorithm is proposed for this purpose. The idea behind the development of this algorithm is a reformulation of the power allocation game as a updating process where countries make myopic transitions from one power allocation matrix to another, where the discretization of the infinite strategy space is key.








In the sequel the steps of the algorithm will be consecutively explained: 

\subsubsection*{Input of the Algorithm.} The input of the algorithm consists of the countries set, $\mathbf{n}$, the total power vector $p$, the friend set, the adversary set and the preference order of each country $i$, $\mathcal{A}_{i}, \mathcal{F}_{i}, \preccurlyeq_{i}; i \in \mathbf{n}$.


\subsubsection*{Initialization.} Initialize the equilibrium set to be an empty set $\mathcal{U}^*_{0} = \emptyset,$ and an initial power allocation matrice set $\tilde{\mathcal{U}}$ whose cardinality is $q > 0$ to be a randomly sampled subset from $\mathcal{U}$.

\subsubsection*{Update Process}   By \emph{update process} is meant that starting from an initial power allocation matrix, countries update their own strategies (asynchronously or synchronously) assuming the strategies of the others to be fixed to maximize their utility until a certain number of rounds ($T > 0$ rounds) or when the countries have reached a Nash equilibrium. There are $q$ updating processes in total, one for each initial power allocation matrix in $\tilde{\mathcal{U}}$.

A suitable utility function that satisfies the two preference axioms can be assumed for countries' update process starting from a sampled power allocation matrix. An example of utility functions that will be used in the simulations later is $$f_{i}(U) = \\\begin{cases}
t_{ii}(0) & x_{i}(U) = \text{unsafe} \\
\sum_{j \in \mathcal{F}^{1}_{i} \cup \mathcal{A}^{1}_{i}}t_{ij}(1) & x_{i}(U) \in \{\text{safe, precarious}\}\\
 \end{cases}$$ where $t_{ij}(0)$ or $t_{ij}(1)$ are defined to be $i$'s pairwise utilities from the relation with $j$ depending on whether $j$ is a friend or an adversary as well as $j$'s state predicted by $U$ (\cite{paradox}). 
 
 For instance, $t_{ii}(0)$ is both the total utility of $i$ and the pairwise utility from itself when it has not survived (which can be simplified to be $0$); $t_{ii}(1) > 0$ is its pairwise utility from itself when it has. $\mathcal{F}^{1}_{i}$ denotes the set of $i$'s friends who are safe/precarious, and $\mathcal{A}^{1}_{i}$ denotes the set of $i$'s adversaries who are unsafe/precarious. The pairwise utilities are also a proxy of countries' \emph{relation importance} with every friend and adversary. 

Starting from the $h$-th ($0 \leq h \leq q$) initial power allocation matrix, $U(0)$, Country $i$ at step $t$ of the updating process adjusts its own strategy to maximize its utility assuming the strategies of others to be fixed. $$u_{i}(t) = \text{argmax}_{u_{i}(t)} f_{i}(u_1(t-1), \ldots, u_i(t), \ldots, u_n(t-1)).$$

Store $U(T)$ or the equilibrium $U^*$ into the equilibrium set $\mathcal{U}^*_{h-1}$, $$\mathscr{U}^{*}_{h} = \mathscr{U}^{*}(h-1) \cup U(T)~\text{or}~ \mathscr{U}^{*}_{h}  = \mathscr{U}^{*}_{h-1} \cup U^{*}. $$ (When $T$ is large enough, $U(T)$ still provides insightful information about countries' power allocation even if it might not be an equilibrium.)

Go to the $(h+1)$-th initial power allocation matrix. Start the updating process starting from this new matrix. At the end of the $q$-th updating process, the equilibrium set $\mathcal{U}^*_{q}$ is generated. 

\subsubsection*{Generate Equilibrium Equivalence Classes} Partition the equilibrium set into equilibrium equivalence classes. The cardinalities of each equilibrium equivalence class can then be used for calculating the likelihood for each class.

		

\subsubsection*{Case Study: North Korea-China-US in 2017-2018}

The execution of this algorithm will obviously use simulations. The current Nuclear crisis that involves the United States (1), China (2), Japan (3), South Korea (4), Russia (5) and North Korea (6) is examined. 

The steps of the simulations are as follows. First, obtain the power indices of these countries, for which the ``composite index for nations' capabilities''(\cite{singer1972capability}) will be used. As of 2012, these six countries' CINC indices are: $$[0.139, 0.218, 0.035, 0.023, 0.040, 0.013],$$ which will be used next. Second, randomize the symmetric relations among these countries into ``friend'' (1), ``adversary'' (-1), or ``null'' (0) and the relation importances between 0 and 1. (The importances for ``null'' relations are by default 0.) Third, for each power allocation game assuming each relation configuration, compute the pure strategy Nash equilibrium set (where 10,000 initial power allocation matrices will be sampled). The previous utility function is to be used, where the sampled relation importances of each country $i$ become its pairwise utilities $t_{ij}(1), j \in \mathbf{n}$; in addition, $t_{ii}(0) = 0$. For simplicity, the relation importances between $i$ and $j$ are assumed to be symmetric in this paper. Lastly, compare the likelihood of survival for each country across different cases.

\begin{center}
\tiny
 \begin{tabular}{||c c c c ||} 
 \hline
First Country & Second Country & Relation Type & Relation Importance \\ [0.5ex] 
 \hline\hline
1	& 1	 & 1	&1\\
\hline
2	& 2	 & 1	&1\\
\hline
3	& 3	& 1	   &1\\
\hline
4	& 4	 & 1	&1\\
\hline
5	& 5	 & 1	&1\\
\hline
6	& 6	 & 1	&1\\
\hline
1	&2	& -1 & 0.3016625\\
\hline
1	&3	&-1 & 0.2175963\\
\hline
1	&4	&1 &0.2161213 \\
\hline
1	&5	&0 &0 \\
\hline
1	&6	&1 &0.3415068 \\
\hline
2	&3	&1 &0.3632781 \\
\hline
2	&4	&0 & 0\\
\hline
2	&5	&0 & 0 \\
\hline
2	&6	&-1 &0.1707308\\
\hline
3	&4	&0 & 0.2319436\\
\hline
3	&5	&1 &0.1440208\\
\hline
3	&6	&0 &0  \\
\hline
4	&5	&0 &0\\
\hline
4	&6	&-1 &0.1355878 \\
\hline
5	&6	&-1 &0.2719213\\
 \hline\hline
\end{tabular}
\end{center}

In the first case, the likelihoods of each of the six countries' survival (characterized by the percentage of the matrices in $\mathcal{U}^*_q$ predicting the country's survival) are: 
$[100\%~100\% ~100\% ~100\% ~100\% ~99.71\%]$. But in most of these cases, most of the countries are in the precarious states. The likelihoods of the six countries' being safe (not precarious) are: $$[0\% ~98.5\% ~6.18\% ~30.54\% ~ 81.63\% ~ 0\%].$$

\begin{center}
\tiny
 \begin{tabular}{||c c c c ||} 
 \hline
First Country & Second Country & Relation Type & Relation Importance \\ [0.5ex] 
 \hline\hline
1	& 1	 & 1	&1\\
\hline
2	& 2	 & 1	&1\\
\hline
3	& 3	& 1	   &1\\
\hline
4	& 4	 & 1	&1\\
\hline
5	& 5	 & 1	&1\\
\hline
6	& 6	 & 1	&1\\
\hline
1	&2	&-1	&0.180719673\\
\hline
1	&3	&0	&0 \\
\hline
1	&4	&1	&0.174140854\\
\hline
1	&5	&1	&0.474582374\\
\hline
1	&6	&0	&0 \\
\hline
2	&3	&-1	&0.164999116\\
\hline
2	&4	&-1	&0.43771212\\
\hline
2	&5	&-1	&0.158066073\\
\hline
2	&6	&-1	&0.137276616\\
\hline
3	&4	&-1	&0.277800827\\
\hline
3	&5	&0	&0 \\
\hline
3	&6	&-1	&0.154644001\\
\hline
4	&5	&0	&0 \\
\hline
4	&6	&1	&0.485971539\\
\hline
5	&6	&0	&0\\
 \hline\hline
\end{tabular}
\end{center}

In the second case, the likelihoods of each of the six countries' survival are: $$[100\% ~100\% ~96.26\% ~97.63\% ~ 100\% ~ 88.29\%].$$ The likelihoods of the six countries' being safe (not precarious) are: $$[8.09\% ~54.07\% ~9.39\% ~ 0\% ~30.92\% ~6.26\%].$$


\section{Conclusion}
This paper examines how to make real world predictions based on the pure strategy Nash equilibrium set of the power allocation game, where the key task is to generate the equilibrium set efficiently. Both methods of a more analytical flavor and simulations are discussed. 

In future work, it should be promising to perform more predictions of countries' power allocation outcomes assuming different networked international environments using simulations for scenarios that have or have not taken place. The predictions will hopefully be a basis for further insights on possible kinds of changes to the environment that could shift the then situation in a country's own favor.

\bibliographystyle{ifacconf}

\end{document}